\documentclass[11pt]{article}
\usepackage{amsmath}
\usepackage{amssymb}
\usepackage{amsthm}
\usepackage{amsfonts}
\usepackage{comment}
\usepackage{setspace}
\usepackage{color}
\usepackage{mathrsfs}
\usepackage{nccmath}
\usepackage{hyperref}
\usepackage{braket}
\usepackage{graphicx}
\usepackage[subrefformat=parens]{subcaption}
\usepackage{cite}

\usepackage[stable]{footmisc}

\makeatletter

\@addtoreset{equation}{section}
\makeatother

\begin{document}

\pagestyle{empty}

\begin{flushright}
\end{flushright}

\vspace{3cm}

\begin{center}

{\bf\LARGE  
Bound on the Lyapunov exponent\\in Kerr-Newman black holes via a charged particle
}
\\

\vspace*{1.5cm}
{\large 
Naoto Kan\footnote{naotokan000@gmail.com} and Bogeun Gwak\footnote{rasenis@dgu.ac.kr}
} \\
\vspace*{0.5cm}

{\it 
Division of Physics and Semiconductor Science, Dongguk University, Seoul 04620,\\Republic of Korea
}

\end{center}

\vspace*{1.0cm}

\begin{abstract}
{We investigate the conjecture on the upper bound of the Lyapunov exponent for the chaotic motion of a charged particle around a Kerr-Newman black hole. The Lyapunov exponent is closely associated with the maximum of the effective potential with respect to the particle. We show that when the angular momenta of the black hole and particle are considered, the Lyapunov exponent can exceed the conjectured upper bound. This is because the angular momenta change the effective potential and increase the magnitude of the chaotic behavior of the particle. Furthermore, the location of the maximum is also related to the value of the Lyapunov exponent and the extremal and non-extremal states of the black hole.     
}
\end{abstract}

\newpage
\baselineskip=18pt
\setcounter{page}{2}
\pagestyle{plain}
\baselineskip=18pt
\pagestyle{plain}

\setcounter{footnote}{0}

\section{Introduction}
\noindent 
The anti-de Sitter/conformal field theory (AdS/CFT) correspondence \cite{Maldacena:1997re} is one of the most remarkable discoveries in quantum gravity. 
The AdS/CFT correspondence first proposed by Maldacena was obtained from the limit of an $N$ D3-branes system.
This system is described by the D3-brane solution of Type IIB supergravity theory from the viewpoint of gravity.
Supergravity is the low-energy effective theory of Type IIB string theory, which is valid in the large-$N$ limit.
In the near-horizon region, this geometry becomes ${\rm AdS}_5\times S^5$.
This region corresponds to the weak coupling region of supergravity.
From the viewpoint of quantum field theories, the effective theory on $N$ D3-branes is described by ${\cal N}=4$ $U(N)$ supersymmetric Yang-Mills theory.
The weak coupling region on the gravity side corresponds to the strong coupling region on the field theory side.
We can calculate the physical quantities of strongly coupled field theories from weakly coupled gravity theories according to the AdS/CFT duality \cite{Gubser:1998bc,Witten:1998qj}. 
For example, the correlation functions of a scalar primary operator in a CFT are encoded in the scatterings of a dual scalar field in AdS spacetime.
In the AdS/CFT correspondence, a CFT with a finite temperature is associated with gravity with an AdS black hole \cite{Witten:1998qj,Witten:1998zw}.
Here, the Hawking temperature is coincident with that of the CFT.

In classical systems, when trajectories of dynamics sensitively depend on the initial conditions, the system is called chaotic, and this phenomenon is called the butterfly effect.
In chaotic systems, close trajectories exponentially diverge.
The sensitivity of classical systems to the initial condition is measured by the Lyapunov exponent.
When the Lyapunov exponent is positive, the system is chaotic.
A chaotic quantum system is characterized by the quantity \cite{Larkin:1969qu}
		\begin{align}
		C(t)=-\braket{[W(t),V(0)]^2}_\beta,
	\end{align}
where $V$ and $W$ are Hermitian operators, and $\braket{\cdots}_\beta={\rm Tr}\, e^{-\beta H}/Z$ is the thermal expectation value.
$\beta$ is the inverse temperature, $\beta=1/T$\footnote{In this study, we assume $k_B=1$.}.
The quantity $C(t)$ includes the out-of-time-ordered correlators (OTOCs) such as $\braket{V(0)W(t)V(0)W(t)}_\beta$.
For early times, the behavior of the quantity $C(t)$ in chaotic systems is typically described by $C(t)\sim e^{2\lambda_L t}$.
We can interpret $\lambda_L$ as the Lyapunov exponent in quantum systems.
In fact, in the semiclassical limit, $\hbar\ll 1$, the Lyapunov exponent $\lambda_L$ measures the sensitivity of the systems to the initial conditions.
This definition of the quantum Lyapunov exponent is well-defined when the collusion time $t_d$ and scrambling time $t_s$ are sufficiently separated, where $t_d\sim \beta$ and $t_s\sim \frac{1}{\lambda_L}\log \frac{1}{\hbar}$.

Recently, Maldacena, Shenker, and Stanford conjectured that the Lyapunov exponent $\lambda_L$ is upper bounded in thermal quantum systems \cite{Maldacena:2015waa}.
They derived the bound by considering shock waves near the horizon of a black hole via the AdS/CFT correspondence \cite{Shenker:2013pqa,Shenker:2013yza}.
According to the conjectured bound, the maximum value of the Lyapunov exponent is proportional to the temperature of the system.
This bound has been studied intensively in the Sachdev-Ye-Kitaev (SYK) model \cite{Sachdev:1992fk,Kitaev:2014hi,Polchinski:2016xgd}, which is the one-dimensional fermionic model with random couplings \cite{Maldacena:2016hyu,Witten:2016iux,Gross:2016kjj}.
The SYK model was also studied via the AdS/CFT correspondence \cite{Kitaev:2015as,Maldacena:2016upp,Kitaev:2017awl}.
For example, the Lyapunov exponent in the SYK model can be calculated for the Jackiw-Teitelboim (JT) gravity \cite{Jackiw:1984je,Teitelboim:1983ux,Maldacena:2016upp}.

The Lyapunov exponent has also been extensively investigated from the viewpoint of black hole systems with a probe particle \cite{Dettmann:1994dj,Suzuki:1996gm,Suzuki:1999si,Dalui:2018qqv,Li:2018wtz,Lukes-Gerakopoulos:2016xoc,Verhaaren:2009md,Han:2008zzf,Cornish:2001jy,Cardoso:2008bp,Pradhan:2012rkk,Pradhan:2012qf,Pradhan:2013bli,Pradhan:2014tva,Jawad:2016kgt,Jai-akson:2017ldo,Chen:2016tmr} (see also \cite{Yurtsever:1994yb,Barrow:1981sx,Bombelli:1991eg,Letelier:1997uv,deMoura:1999wf,Yi:2020shw,Zelenka:2019nyp,Mondal:2021exj,Lukes-Gerakopoulos:2016udm,Setare:2010zd,Liu:2017fjx}) and of the AdS/CFT correspondence \cite{PandoZayas:2010xpn,Basu:2012ae,Hashimoto:2016wme,Nunez:2018ags,Hashimoto:2018fkb,Akutagawa:2018yoe,Akutagawa:2019awh,Cubrovic:2019qee}.
According to the AdS/CFT duality, a probe particle near the event horizon of a black hole is dual to some operator on the field theory side.
The authors of \cite{Hashimoto:2016dfz} studied the bound on the Lyapunov exponent of such a probe particle.
They assumed a static and spherically symmetric black hole and introduced external forces such as the electrostatic force and the scalar force.
The effective potential of the particle around the local maximum is described by the inverse harmonic potential, which causes the butterfly effect (e.g., see \cite{Morita:2021syq,Morita:2021mfi,Hashimoto:2016wme,Bhattacharyya:2020art}).
They calculated the maximum value of the Lyapunov exponent of the probe particle.
In these cases, they found that the maximum value coincides with the upper bound given by Maldacena et al.
They also considered the particle with higher spin forces as the external force.
In that case, they found that the bound on the Lyapunov exponent can be violated.
In \cite{Zhao:2018wkl}, the Lyapunov exponents were calculated for many concrete black holes, e.g., asymptotically flat, AdS, and de Sitter (dS) Reissner-Nordstr\"om (RN) black holes.
The authors of \cite{Zhao:2018wkl} found that for an asymptotically dS black hole, the bound is satisfied only for the near-horizon region.
In other words, the bound is violated for the dS black hole if the local maximum is located away from the horizon.
In \cite{Lei:2020clg}, the Lyapunov exponent for a particle around a black hole with quasi-topological electromagnetism was considered.
The violation of the bound was also found.

In this work, we investigate the bound of the Lyapunov exponent on the particle motion in Kerr-Newman (KN) black holes. The Lyapunov exponent was expected to be bounded under the surface gravity in black hole systems, but the system with an electromagnetic field could exceed the bound beyond the surface gravity\cite{Hashimoto:2016dfz}. Here, we consider the KN black hole and a particle with an angular momentum to introduce the centrifugal force. Hence, we generalize the analysis of the Lyapunov exponent to black hole systems with centrifugal force as an effective force. Furthermore, the investigated system includes the angular momenta of the black hole and particle and therefore has a richer structure and is more  complicated than those of previous studies such as \cite{Hashimoto:2016dfz,Zhao:2018wkl,Lei:2020clg}, which motivated our work. We show that {\it the angular momenta of the KN black hole and particle play a significant role in the violation of the bound.} This implies that the change in the gravitational potential owing to the effective force can affect the chaotic behavior of the system. The angular momentum of the black hole also makes a huge difference in the case of RN black holes because its value becomes a constraint on the location of the maximum in the gravitational potential.

The remainder of this paper is organized as follows:
In Sec. \ref{sec:review}, we review the physics of the KN black hole, the chaotic behavior of the inverse harmonic oscillators, and the bound on the Lyapunov exponent.
In particular, we review that the upper bound of the Lyapunov exponent is given by the surface gravity in the black hole systems.
In Sec. \ref{sec:particle_motion}, we consider the motion of a charged particle around a KN black hole.
We calculate the effective potential of the particle, and using this, we obtain the upper bound on the Lyapunov exponent of the particle.
In Sec. \ref{sec:analysis_of_Lyapunov_exponent}, we evaluate the bound on the Lyapunov exponent.
This section consists of three subsections, in which we consider the KN, Kerr, and RN black holes, respectively.
We also consider the near-horizon region in each subsection.
Finally, in Sec. \ref{sec:conclusions}, we summarize our conclusions.

\section{Review}
\label{sec:review}
\subsection{The Kerr-Newman black hole}
\noindent
The KN black hole is the solution to the Einstein-Maxwell theory of gravity. It demonstrates a black hole with spinning angular momentum and electric charge. In Boyer-Lindquist coordinates, the metric is given as
	\begin{align}
		ds^2=-\frac{\Delta}{\rho^2}\left(dt-a \sin^2\theta\, d\phi\right)^2+\frac{\sin^2\theta}{\rho^2}\left(adt-(r^2+a^2)d\phi\right)^2+\frac{\rho^2}{\Delta}dr^2+\rho^2d\theta^2,
	\label{eq:metric_Kerr_Newman}
	\end{align}
where $\Delta=r^2-2Mr+a^2+Q^2$ and $\rho^2=r^2+a^2\cos^2\theta$. Here, $M$ and $Q$ are the mass and electric charge of the black hole, respectively. The angular momentum of the KN black hole is denoted by $J$, and the spin parameter is defined as $a=J/M$. The gauge potential is 
	\begin{align}
		A=\frac{Qr}{\rho^2}dt-\frac{aQr\sin^2\theta}{\rho^2}d\phi.
	\end{align}
The locations of the horizons are given by
	\begin{align}
		r_\pm=M\pm \sqrt{M^2-(a^2+Q^2)},
	\end{align}
where $r_+$ is the event horizon, and $r_-$ is the Cauchy horizon. At the outer horizon, the angular velocity, surface area, and surface gravity are
\begin{align}
    \Omega_+=\frac{a}{r_+^2+a^2},\quad \kappa=\frac{r_+\left(1-\frac{a^2}{r_+^2}\right)}{2(r_+^2+a^2)},\quad A_+=4\pi(r_+^2+a^2).
\end{align}
The surface gravity is crucial in our analysis of the Lyapunov exponent, which will be introduced in the following section.
\subsection{Inverse harmonic oscillators and the bound on the Lyapunov exponent}
\label{subsec:review_bound}
\noindent
Inverse harmonic oscillators are known to be associated with the butterfly effect in chaos.
The Lyapunov exponent is a quantity that measures the sensitivity of a dynamic system to the initial condition.
Recently, the bound on the Lyapunov exponent was proposed by Maldacena, Shenker, and Stanford \cite{Maldacena:2015waa} via the AdS/CFT correspondence.
Here, we briefly review the Lyapunov exponent in inverse harmonic oscillators and the bound on the Lyapunov exponent.

The effective motion of a particle around a black hole is described by the inverse harmonic oscillator.
The equation of motion for the inverse harmonic oscillator is given by
	\begin{align}
		m\ddot{x}-kx=0,
	\label{eq:eom_inverse_harmonic_oscillator}
	\end{align}
where $m$ is the mass of a particle and $k>0$.
The solution to \eqref{eq:eom_inverse_harmonic_oscillator} is given by
	\begin{align}
		x(t)=C_1e^{-\omega t}+C_2 e^{\omega t},
	\end{align}
where $C_1$ and $C_2$ are integration constants, and $\omega=\sqrt{k/m}$.
In classical chaotic systems, the divergence of close trajectories increases exponentially,
	\begin{align}
		\frac{\partial x(t)}{\partial x(0)}=\left\{x(t),p(0)\right\}_{\rm PB}\sim e^{\lambda t},
	\label{eq:def_Lyapunov_exponent}
	\end{align} 
	where $\{\cdots\}_{\rm PB}$ is the Poisson bracket, and $\lambda$ is the Lyapunov exponent, which measures the sensitivity to the initial condition.
The Lyapunov exponent of this system is given by
	\begin{align}
		\lambda=\omega=\sqrt{\frac{k}{m}}.
	\label{eq:Lyapunov_inverse_harmonic_oscillator}
	\end{align}

In thermal quantum chaotic systems, the Lyapunov exponent is conjectured to be bounded. 
The Lyapunov exponent $\lambda_L$ in a quantum system is defined using OTOCs, which is expected as a quantum version of the classical Lyapunov exponent defined in \eqref{eq:def_Lyapunov_exponent}.
The bound is given by
	\begin{align}
		\lambda_{\rm L}\le \frac{2\pi T}{\hbar},
	\label{eq:MSS_bound}
	\end{align}
where $T$ is the temperature of the system.
This bound was originally derived by AdS/CFT correspondence \cite{Maldacena:2015waa}.
We note that \eqref{eq:MSS_bound} is the bound for quantum systems without gravity.

We apply the bound \eqref{eq:MSS_bound} to a system of a particle around a black hole according to the perspective of holography.
The temperature of the system is given by the Hawking temperature $T_{\rm BH}$, and we obtain
	\begin{align}
		\lambda\le \frac{2\pi T_{\rm BH}}{\hbar}=\kappa,
	\label{eq:black_hole_bound}
	\end{align}
where $\kappa$ is the surface gravity of the black hole.
In particular, if the system is chaotic (i.e., $\lambda>0$), we can square both sides of \eqref{eq:black_hole_bound}:
	\begin{align}
		\lambda^2\le \kappa^2.
	\label{eq:black_hole_bound_square}
	\end{align}
In the next section, we consider the motion of a probe particle around a KN black hole in four dimensions.
The inverse harmonic potential will appear as an effective potential for the radial direction of the particle.
Then \eqref{eq:Lyapunov_inverse_harmonic_oscillator} provides a maximum value of the Lyapunov exponent.
We simply refer to this maximum value as the Lyapunov exponent $\lambda$  in the rest of this paper.

\section{The Lyapunov exponent of the particle with the static gauge}
\label{sec:particle_motion}
\noindent
We consider the motion of a probe particle around a KN black hole with the static gauge.
When the particle is at a local maximum of an effective potential, the motion of the particle is described by an inverse harmonic oscillator.

We start with the Polyakov-type action, which is identical to the Nambu-Goto-type action.
The action of a particle with charge $q$ and mass $m$ in the curved space is given by
	\begin{align}
		S=\int ds \left[\frac{1}{2e(X(s))}g_{\mu\nu}(X(s))\dot{X}^\mu(s) \dot{X}^\nu(s)-\frac{e(X(s))}{2}m^2-qA_\mu(X(s))\dot{X}^\mu(s)\right],
	\end{align}
where $e$ is an auxiliary field, and $s$ parametrizes the geodesic of the particle.
The last term on the right-hand side is the interaction term between the particle and the electromagnetic force.
We work on a static gauge, $X^0=s$.
Then the action in the KN metric \eqref{eq:metric_Kerr_Newman} is
	\begin{align}
		S=&\int ds \bigg[\frac{1}{2e}\bigg(-\frac{1}{\rho^2}\left(\Delta-a^2\sin^2\theta\right)+\frac{\rho^2}{\Delta}\dot{r}^2+\rho^2\dot{\theta}^2+\frac{2a\sin^2\theta}{\rho^2}\left(\Delta-(r^2+a^2)\right)\dot{\phi} \nonumber \\
		&\qquad \qquad+\frac{\sin^2\theta}{\rho^2}\left((r^2+a^2)^2-\Delta a^2\sin^2\theta\right)\dot{\phi}^2\bigg)-\frac{e}{2}m^2-q\frac{Qr}{\rho^2}+q\frac{aQr\sin^2\theta}{\rho^2}\dot{\phi}\bigg].
	\end{align}
We focus on equatorial motion of the particle.
Assuming $\theta=\pi/2$, we find that 
	\begin{align}
		S=&\int ds\bigg[\frac{1}{2e}\bigg(-\frac{1}{r^2}\left(\Delta-a^2\right)+\frac{r^2}{\Delta}\dot{r}^2+\frac{2a}{r^2}\left(\Delta-(r^2+a^2)\right)\dot{\phi} \nonumber \\
		&\qquad \qquad +\frac{1}{r^2}\left((r^2+a^2)-\Delta a^2\right)\dot{\phi}^2\bigg) -\frac{e}{2}m^2-q\frac{Q}{r}+q\frac{a Q}{r}\dot{\phi}\bigg].
	\label{eq:original_action}
	\end{align}
	
The action \eqref{eq:original_action} depends on $\dot{\phi}$, but not on $\phi$.
The action is invariant under translation for $\phi$.
In order to obtain an effective action, we calculate an angular momentum:
	\begin{align}
		L=\frac{\partial {\cal L}}{\partial\dot{\phi}} 	=\frac{a}{er^2}\left(\Delta-(r^2+a^2)\right)+\frac{1}{er^2}\left((r^2+a^2)^2-\Delta a^2\right)\dot{\phi}+q\frac{aQ}{r},
	\label{eq:angular_momentum}
	\end{align}
where the Lagrangian is defined as $S=\int ds\ {\cal L}$.
Furthermore, to erase the auxiliary field $e$ from the Lagrangian, we use the equation of motion for $e$,
	\begin{align}
		-\frac{1}{r^2}\left(\Delta-a^2\right)+\frac{2a}{r^2}\left(\Delta-(r^2+a^2)\right)\dot{\phi}+\frac{1}{r^2}\left((r^2+a^2)^2-\Delta a^2\right)\dot{\phi}^2+\frac{r^2}{\Delta}\dot{r}^2=-e^2m^2.
	\label{eq:eom_e}
	\end{align}
Using \eqref{eq:angular_momentum} and \eqref{eq:eom_e}, we obtain the effective Lagrangian,
	\begin{align}
		{\cal L}_{\rm eff}=&{\cal L}-L\dot{\phi} \nonumber \\
		=&\, e\left[-\frac{m^2}{2}-\frac{1}{2}\frac{(aqQ-Lr)^2}{(r^2+a^2)^2-\Delta a^2}\right]-\frac{1}{2e}\left[\frac{\Delta-a^2}{r^2}-\frac{r^2}{\Delta}+\frac{a^2\left(\Delta-(r^2+a^2)\right)^2}{r^2\left((r^2+a^2)^2-\Delta a^2\right)}\right] \nonumber \\
		&-\frac{a(aqQ-Lr)\left(\Delta-(r^2+a^2)\right)}{r\left((r^2+a^2)^2-\Delta a^2\right)}-\frac{qQ}{r}
	\label{eq:eff_Lagrangian_relativistic}
	\end{align}
with
	\begin{align}
		e=r\sqrt{\frac{\Delta^2-\left((r^2+a^2)^2-\Delta a^2\right)\dot{r}^2}{\alpha(r)}},
	\end{align}
where 
	\begin{align}
		\alpha(r)=\Delta\left[m^2\left((r^2+a^2)^2-\Delta a^2\right)+(aqQ-Lr)^2\right].
	\end{align}
Here $\alpha(r)$ is positive when $r>r_+$.
	
We focus on the motion around the local maximum of a potential around the local maximum; the initial velocity of the particle is quite slow.
Thus, the motion of the particle is described by the non-relativistic limit for the $r$-direction by taking $\dot{r}\ll 1$.
Then, we find the effective Lagrangian for the non-relativistic particle from \eqref{eq:eff_Lagrangian_relativistic}:
	\begin{align}
		{\cal L}_{\rm eff}=\frac{1}{2}K(r)\dot{r}^2-V_{\rm eff}(r)+O(\dot{r}^4),
	\label{eq:effective_Lagrangian_non_relativistic}
	\end{align}
where
	\begin{align}
		K(r)=\frac{r\sqrt{\alpha(r)}}{\Delta^2},
	\label{eq:def_kinetic}
	\end{align}
and 
	\begin{align}
		V_{\rm eff}(r)=\frac{1}{(r^2+a^2)^2-\Delta a^2}\left(r\sqrt{\alpha(r)}-aL\Delta+(r^2+a^2)(qQr+aL)\right).
	\label{eq:effective_potential}
	\end{align}

Let us consider the motion of the particle around the local maximum.
The position of the local extrema $r_0$ is obtained by solving $V_{\rm eff}'(r)=0$.
We expand the effective Lagrangian \eqref{eq:effective_Lagrangian_non_relativistic} around $r=r_0$.
With a small perturbation $r(s)=r_0+\epsilon(s)$, the effective Lagrangian becomes
	\begin{align}
		{\cal L}_{\rm eff}=\frac{1}{2}K(r_0)\left(\dot{\epsilon}^2+\lambda^2\epsilon^2\right),
	\label{eq:eff_Lagrangian_inverse_oscillator}
	\end{align}
where we neglect constant terms and higher-order terms.
Especially, the coefficient of $\epsilon^2$ is the Lyapunov exponent:
	\begin{align}
		\lambda^2=-\frac{V_{\rm eff}''(r_0)}{K(r_0)},
	\label{eq:Lyapunov_exponent}
	\end{align}
where the second derivative of the effective potential is given by
	\begin{align}
		V''_{\rm eff}(r)=&\frac{1}{4\left((r^2+a^2)^2-\Delta a^2\right)\alpha^{3/2}}\left(24 q Q r \alpha^{3/2}-r \alpha'^2+2 \alpha\left(r \alpha''+2 \alpha'\right)\right) \\
		&-\frac{8r \left(a^2+r^2\right)-2a^2 \Delta'}{\left((r^2+a^2)^2-\Delta a^2\right)^2} \left(a^2 q Q-a L \Delta'+2 a L r+3 q Q r^2+\frac{r \alpha'}{2\sqrt{\alpha}}+\sqrt{\alpha}\right) \nonumber \\
		&-\frac{2\left(\left(a^2+r^2\right) (a L+q Q r)-a L \Delta+r \sqrt{\alpha}\right)}{\left((r^2+a^2)^2-\Delta a^2\right)^3} \nonumber \\
		&\qquad\times\left(\left(a^2+6 r^2\right) \left(\left(a^2+r^2\right)^2-a^2 \Delta\right)-\left(a^2 \Delta'-4 r \left(a^2+r^2\right)\right)^2\right) \nonumber \\
	\label{eq:second_derivative_eff_potential}
	\end{align}
Note that when $V''_{\rm eff}(r_0)<0$, i.e., $\lambda^2>0$, the extremum associated with $r_0$ is the local maximum.
Then, the effective Lagrangian \eqref{eq:eff_Lagrangian_inverse_oscillator} describes the inverse harmonic oscillator, which describes a chaotic system.

Because the position of the local maximum is determined by the equation $V'_{\rm eff}(r)=0$, the Lyapunov exponent depends on the parameters of the black hole and particle.
However, analytically solving $V'_{\rm eff}(r)=0$ is difficult in general; therefore, we perform numerical calculation in the next section.

\section{Analysis of the Lyapunov exponent}
\label{sec:analysis_of_Lyapunov_exponent}
\noindent
Let us analyze the bound \eqref{eq:black_hole_bound_square} for the KN black hole and its two limits: the Kerr and RN black holes.
We now note that the existence of $a$ and $L$ is important because these parameters contribute to the Lyapunov exponent nontrivially.
We complete the analysis of each type of black hole to observe the conditions under which the Lyapunov exponent can violate the bound.

\subsection{The Kerr-Newman black hole}
\noindent
First, we consider the most general case: the KN black hole.
To find the position of the local maximum, we solve the equation $V_{\rm eff}'(r)=0$ numerically.
We can rewrite the bound \eqref{eq:black_hole_bound_square} as
	\begin{align}
		0\le\kappa^2-\lambda^2.
	\end{align}
Thus the sign of $\kappa^2-\lambda^2$ is important.
We investigate whether  the Lyapunov exponent satisfies the bound or not.

\subsubsection{The extremal cases}
\label{sec:Numerical_analysis_for_the_extremal_Kerr-Newman}
\noindent
We consider the extremal KN black hole. 
Because the Hawking temperature for the extremal black hole is zero, the bound is simply given by
	\begin{align}
		0\le\kappa^2-\lambda^2=-\lambda^2.
	\label{eq:bound_extremal_KN}
	\end{align}
As mentioned in the previous section, when there is a local maximum of the effective potential, $\lambda^2>0$, the bound is violated.
Thus, an important point is the existence of the local maximum; the bound is violated if and only if there is a local maximum.

\begin{figure}[h]
	\centering 
		\begin{minipage}[t]{0.4\hsize}
		\centering
			\includegraphics[scale=0.4]{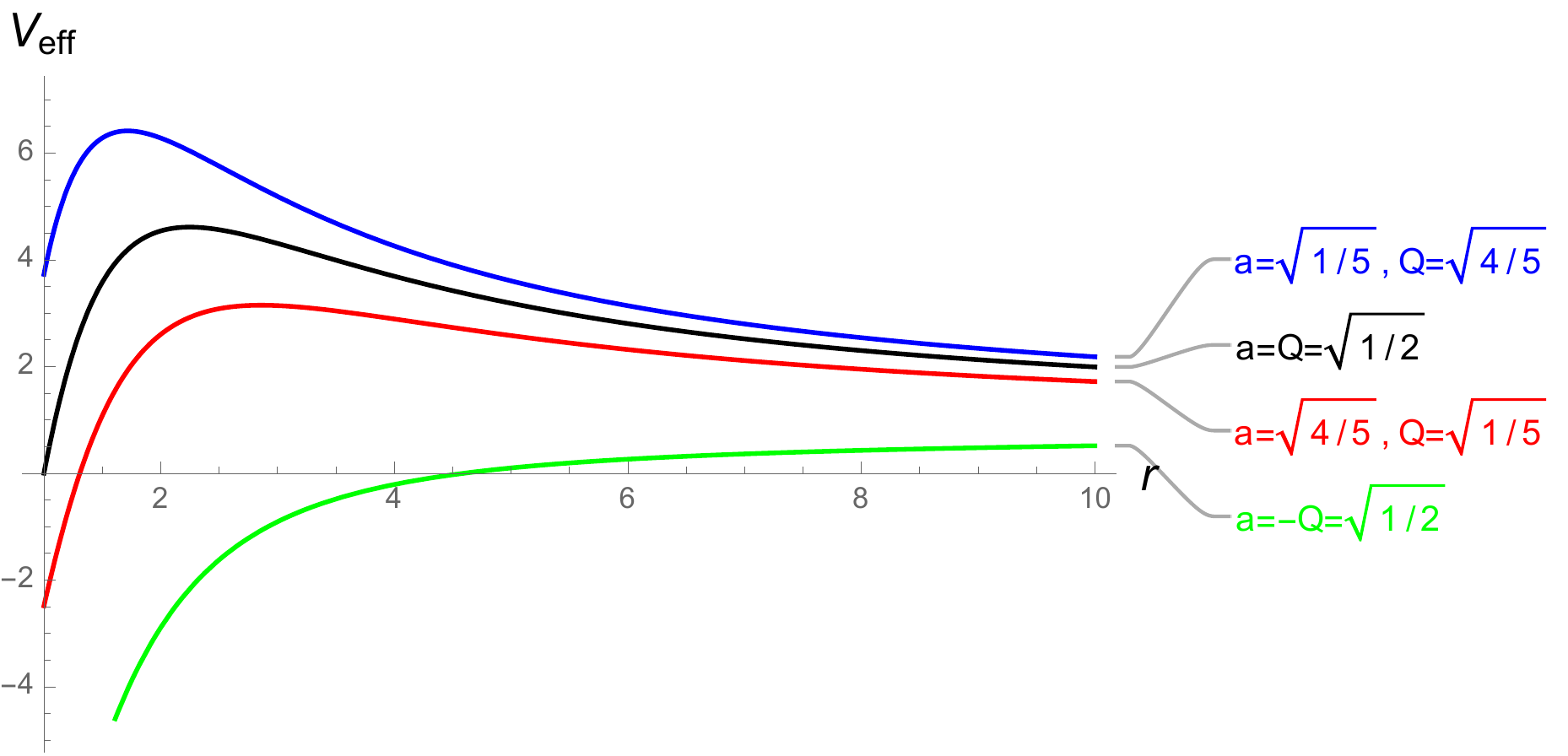}
			\subcaption{}
		\end{minipage}
		\hspace{1.5cm}
		\begin{minipage}[t]{0.4\hsize}
			\includegraphics[scale=0.4]{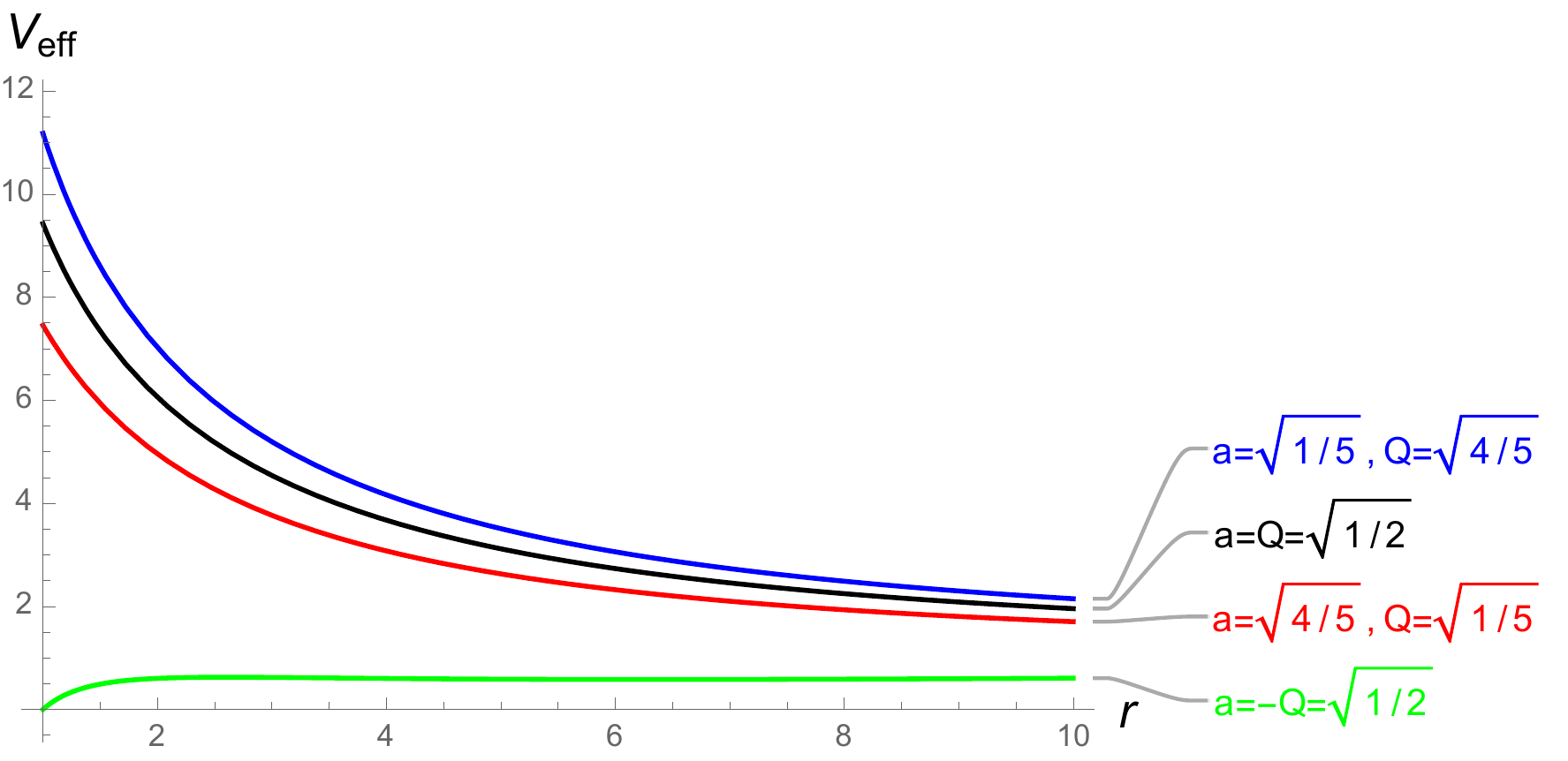}
			\subcaption{}
		\end{minipage}
		\caption{Effective potential of a particle with (a) $L=-10$ and (b) $L=10$ for the extremal Kerr-Newman black hole. \label{fig:eff_potential_exKN_pm10}}
	\end{figure}

We perform numerical calculation to find the position of the local maximum.
We focus on the angular momenta, $a$ and $L$, and the charge of the black hole, $Q$.
For the numerical calculation, we set
	\begin{align}
		M=1,\qquad m=1,\qquad q=10.
	\label{eq:set_parameters}
	\end{align}
Existence of the local maximum depends on the parameters.
We show the effective potential for $L=-10$ and $L=10$ in Fig. \ref{fig:eff_potential_exKN_pm10}.
For example, we see that for $a=Q=\sqrt{1/2}$, we have a local maximum for $L=-10$, but not for $L=10$.
For these parameters, the local maximum exists when $L<(3\sqrt{51}-20)/2\simeq0.712$.

\begin{figure}[h]
	\centering
		\includegraphics	[scale=0.4]{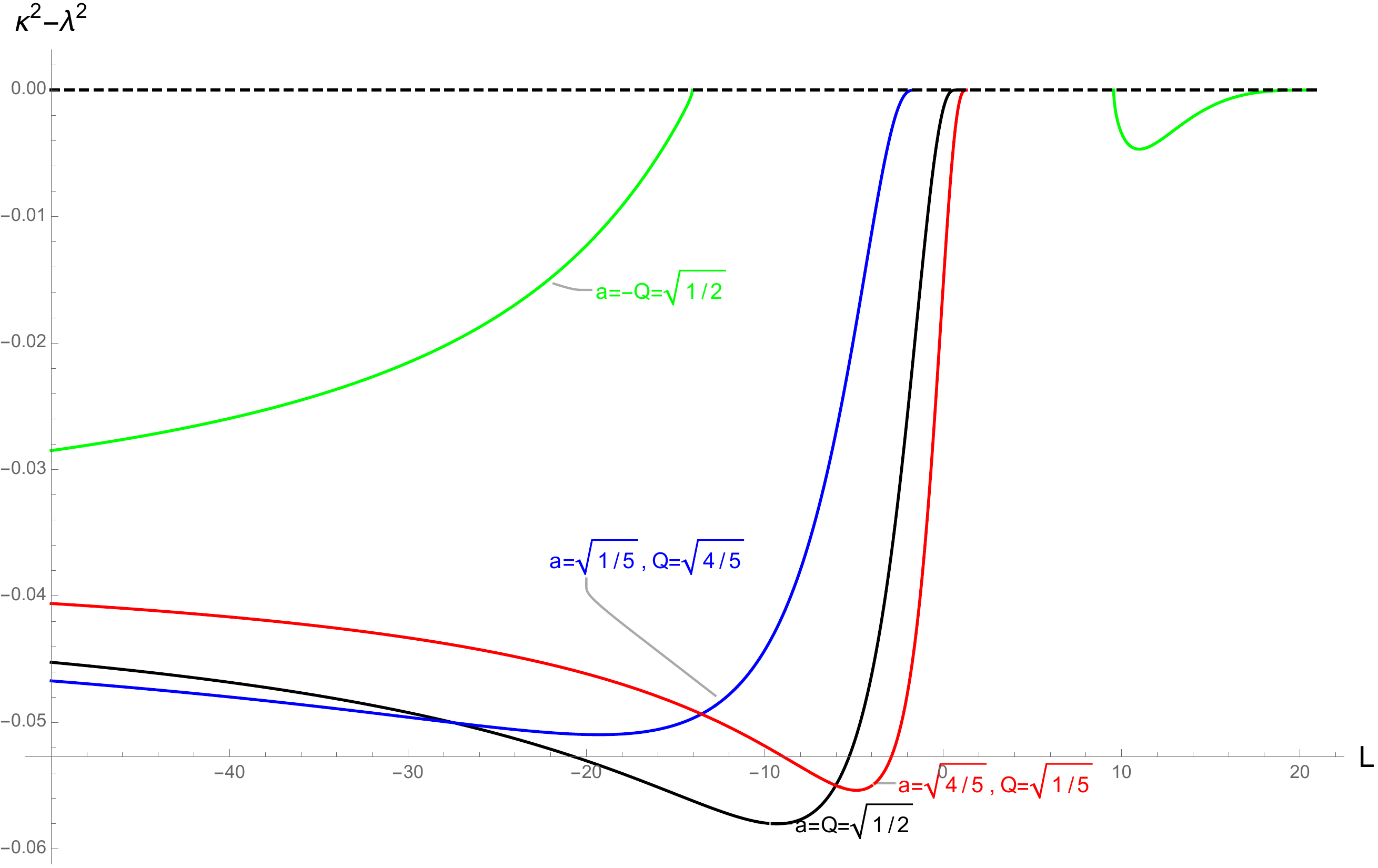}
		\caption{Numerical analysis of $\kappa^2-\lambda^2$ for the extremal Kerr-Newman black hole. The parameters corresponding to each color are as follows: The black line represents $a=Q=\sqrt{1/2}$. The blue line represents $a=\sqrt{1/5}$, $Q=\sqrt{4/5}$. The red line represents $a=\sqrt{4/5}$, $Q=\sqrt{1/5}$. The green line represents $a=-Q=\sqrt{1/2}$. The broken line represents the bound. We find the violation of the bound for all parameter choices. \label{fig:Lyapunov_exKN_All}}
	\end{figure}

The value of $\kappa^2-\lambda^2$ obtained by inserting the numerical results of the positions of the local maxima into \eqref{eq:Lyapunov_exponent} is shown in Fig. \ref{fig:Lyapunov_exKN_All} for the given charge and angular momentum of the black hole.
We see from Fig. \ref{fig:Lyapunov_exKN_All} that for a large negative value of $L$, the Lyapunov exponent becomes large, indicating that the system becomes more chaotic.
For all cases, we find that local maxima exist, and we observe the violation of the bound.

Next, we consider the bound on the Lyapunov exponent in the near-horizon region.
We can easily see from \eqref{eq:Lyapunov_exponent} that the Lyapunov exponent becomes zero if the local maximum is at the horizon, $r_0=r_+$.
Hence, for the extremal black hole, the near-horizon region corresponds to  the vicinity of the boundary of the bound. 
In this region, we can perform algebraic analysis.

To analyze the region, we should find the limits of the parameters corresponding to $r_0\to r_+$.
Supposing $r_0=r_+$, we obtain $V_{\rm eff}'(r_0)=0$ as follows:
	\begin{align}
		a^2 q Q-2 a L r_+-q Qr_+^2+r_+\sqrt{m^2 \left(a^2+r_+^2\right)^2+(a q Q-Lr_+)^2}=0.
	\end{align}
Note that the equation is valid for some finite value of the parameters.
If the parameters are infinite, we should treat the near-horizon limit carefully\footnote{In fact, we will encounter this situation later.}.
	
Introducing a positive small parameter $\epsilon\ (>0)$ as $r_0=r_++\epsilon$, we expand the Lyapunov exponent around $\epsilon=0$. 
This gives
	\begin{align}
		\lambda^2=-\frac{2 \left(\left(a^2+r_+^2\right)\left(r_+\zeta+\xi^2\right)-\xi\chi-4\xi^2r_+^2\right)}{\xi^2r_+\left(a^2+r_+^2\right)^3}\epsilon^3+{\cal O}(\epsilon^4),
	\end{align}
where we define
	\begin{align}
		\xi^2&=a^4 m^2+2 a^2 m^2 r_+^2+a^2 q^2 Q^2-2 a L q Q r_++L^2 r_+^2+m^2r_+^4, \\
		\chi&=a^3 L+2 a^2 q Qr_+-2aLr_+^2-qQr_+^3, \\
		\zeta&=2 a^2 m^2 r_+-a L q Q+L^2r_++2 m^2r_+^3.
	\end{align}
The Lyapunov exponent depends on the distance between the local maximum and the outer horizon.
In particular, when the local maximum is close to the outer horizon, the square of the Lyapunov exponent is proportional to $\epsilon^3$.
Whether the bound is violated or not depends on the parameters.

\subsubsection{The non-extremal cases}
\begin{figure}[h]
	\centering
		\includegraphics[scale=0.4]{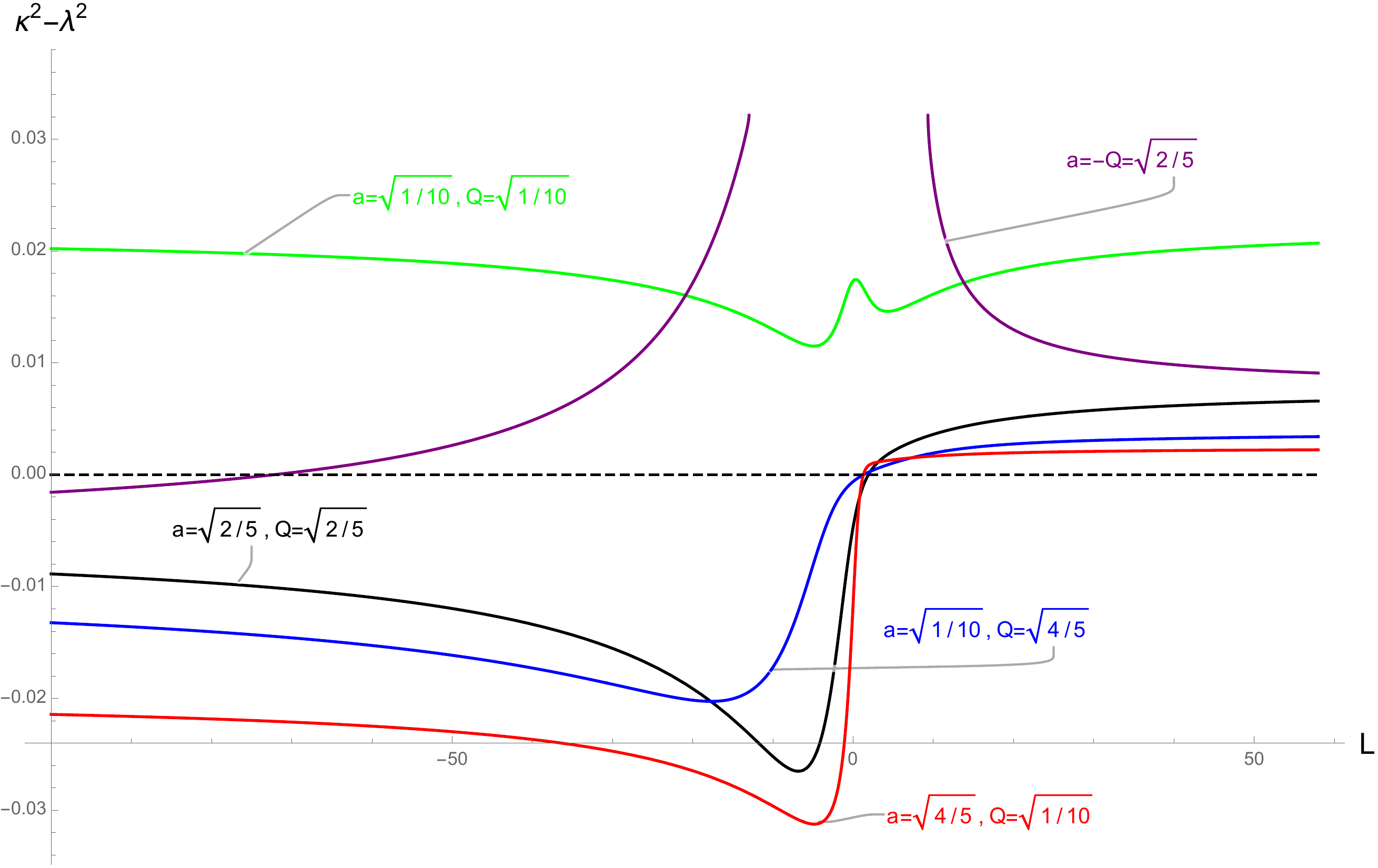}
		\caption{Numerical analysis of $\kappa^2-\lambda^2$ for the non-extremal Kerr-Newman black hole. The parameters corresponding to each color are as follows: The black line represents $a=Q=\sqrt{2/5}$. The blue line represents $a=\sqrt{1/10}$, $Q=\sqrt{4/5}$. The red line represents $a=\sqrt{4/5}$, $Q=\sqrt{1/10}$. The green line represents $a=Q=\sqrt{1/10}$. The purple line represents $a=-Q=\sqrt{2/5}$. The broken line represents the bound. We find the violation of the bound, except for the case of $a=Q=\sqrt{1/10}$.  \label{fig:Lyapunov_nonexKN}}
	\end{figure}

\noindent 
We consider the non-extremal KN black hole.
The bound allows a positive Lyapunov exponent.
For the non-extremal KN black hole, the square of the bound is given by
	\begin{align}
		\lambda^2\le\kappa^2=\left(\frac{\sqrt{M^2-a^2-Q^2}}{2M^2+2M\sqrt{M^2-a^2-Q^2}-Q^2}\right)^2=\frac{(r_+-r_-)^2}{4(r_+^2+a^2)^2}.
	\end{align}
To perform the numerical calculation, we set the parameters as in \eqref{eq:set_parameters}.
The numerical results are shown in Fig. \ref{fig:Lyapunov_nonexKN} for the given black hole parameters.
We see that the bound is violated except for $a=Q=\sqrt{1/10}$ (the green line).
For $a=-Q=\sqrt{2/5}$ (the purple line), the effective potential does not have a local maximum for small $L$.
The violation is observed when the parameters of the black hole are close to the extreme.

Let us consider the case in which the local maximum is near the horizon.
For the non-extremal KN black hole, $V'_{\rm eff}(r_0\to r_+)$ is divergent.
To investigate in more detail, we expand $V'_{\rm eff}(r_++\epsilon)$ around $\epsilon=0$ 
to obtain
	\begin{align}
		V'_{\rm eff}(r_++\epsilon)=&\frac{r_+\sqrt{(r_+-r_-) \left(m^2 \left(a^2+r_+^2\right)^2+(a q Q-Lr_+)^2\right)}}{2\left(a^2+r_+^2\right)^2}\epsilon^{-1/2} \nonumber \\
		&+\frac{a^4qQ-2a^3Lr_++a^2qQr_+(r_+-r_-)+aLr_+^2(r_--3r_+)-qQr_+^4}{\left(a^2+r_+^2\right)^3}+{\cal O}(\epsilon^{1/2}).
	\end{align}
The first term on the right-hand side is nonzero, except for the extremal limit.

To find the near-horizon limit, we assume
	\begin{align}
		q=\frac{Lr_+}{aQ}.
	\label{eq:find_near_horizon_nonexKN}
	\end{align}	
Then $V'_{\rm eff}(r_++\epsilon)=0$ provides
	\begin{align}
		L=\frac{ma\sqrt{r_+-r_-}}{2\sqrt{\epsilon}}+{\cal O}(\epsilon^{1/2}).
	\label{eq:near_horizon_limit_nonexKN}
	\end{align}
Therefore, in the large-$L$ limit (where $q$ is also correspondingly large), the local maximum approaches the horizon.

Expanding the Lyapunov exponent with \eqref{eq:find_near_horizon_nonexKN} and \eqref{eq:near_horizon_limit_nonexKN} around $\epsilon=0$, we obtain
	\begin{align}
		\kappa^2-\lambda^2=\frac{(r_+-r_-)^2 \left(a^2(3r_++r_-)+4 r_+^3\right)}{4\left(a^2+r_+^2\right)^4}\epsilon+{\cal O}(\epsilon^2).
	\end{align}
Thus, the bound is satisfied for the near-horizon limit because the coefficient of the leading term on the right-hand side is positive definite.

\subsection{$Q\to 0$ limit: The Kerr black hole}
In the previous subsection, we discussed the bound for the KN black hole and found the violation.
In this subsection, we focus on the black hole angular momentum.
To identify the effect of the angular momentum, we take the limit of $Q\to0$.
\subsubsection{The extremal cases}
\noindent
We numerically analyze the bound of the Kerr black hole.
To consider the Kerr black hole, we set the charge of the black hole as $Q=0$; the charge $q$ becomes irrelevant.
The Hawking temperature is zero for the extremal black hole; thus, the bound is given by
	\begin{align}
		0\le\kappa^2-\lambda^2=-\lambda^2.
	\end{align}
In the same way as for the extremal KN black hole, the bound is violated when there is a local maximum in the effective potential.

For the extremal Kerr black hole, the parameter associated with the angular momentum of the black hole is given by 
	\begin{align}
		a=\pm M.
	\end{align}
In the effective potential \eqref{eq:effective_potential} with $Q=0$, $a$ appears in the combinations $a^2$ or $aL$; thus, $a=+M$ with $L$ corresponds to $a=-M$ with $-L$.
This implies that the relevant parameter is just $L$.
Because the choice of the mass parameters, $M$ and $m$, is just normalization, our numerical analysis for the extremal Kerr black hole is physically sufficient.

\begin{figure}[h]
	\centering 
		\includegraphics[scale=0.7]{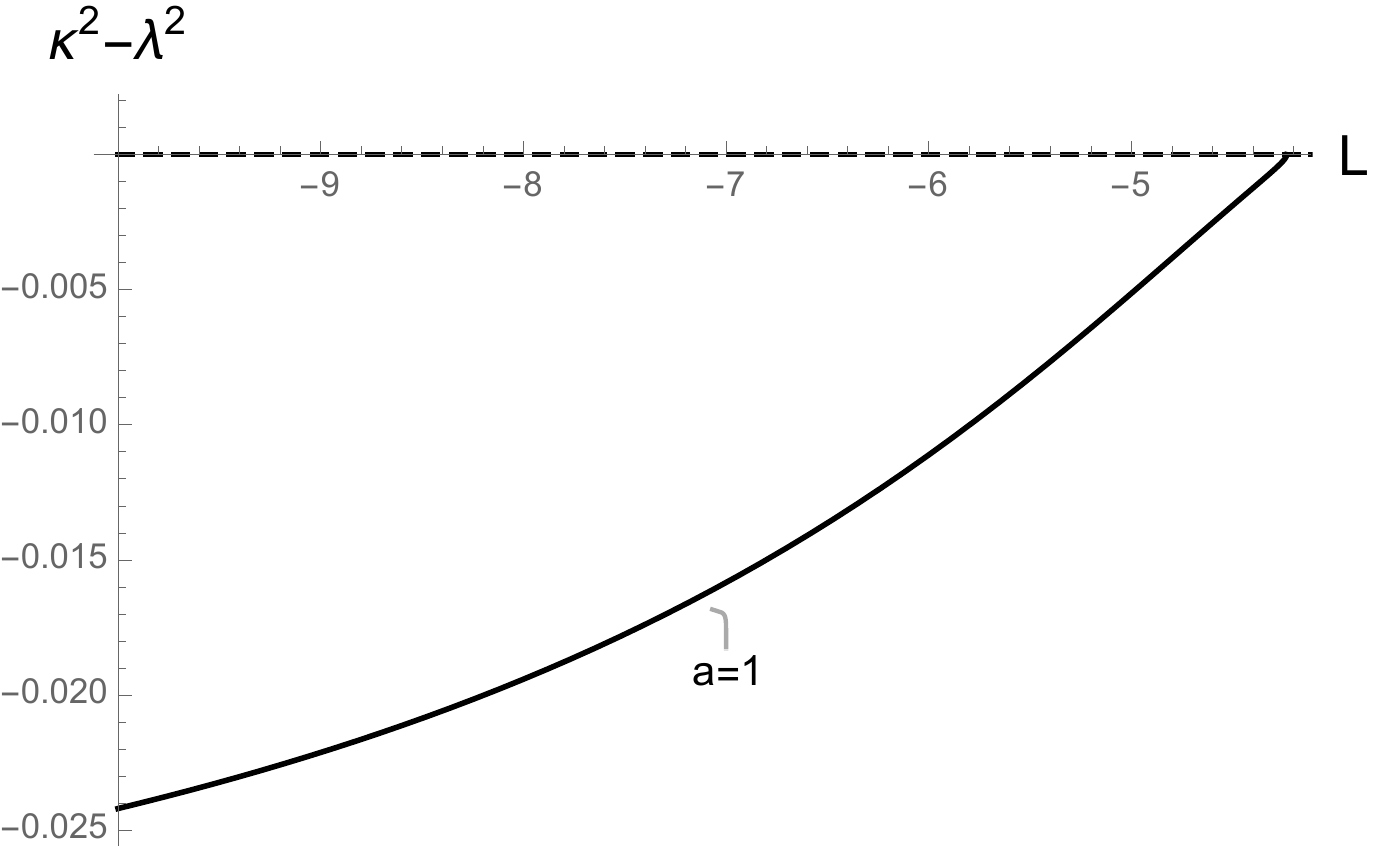}
		\caption{Numerical analysis of $\kappa^2-\lambda^2$ for the extremal Kerr black hole with $a=M=1$. A local maximum exists for $L<-22\sqrt{3}/9$. We see that the bound is violated. \label{fig:Lyapunov_exKerr}}
	\end{figure}

The numerical result is shown in Fig. \ref{fig:Lyapunov_exKerr}.
There is a local maximum for $L<-22\sqrt{3}/9$.
We see from Fig. \ref{fig:Lyapunov_exKerr} that the bound is violated when the local maximum exists.
When the angular momentum $L$ becomes a large negative value, the Lyapunov exponent becomes large, meaning that the system becomes more chaotic.

Let us consider the near-horizon limit of the local maximum. 
We expand $V'_{\rm eff}(r_++\epsilon)$ around $\epsilon=0$ to obtain
	\begin{align}
		V'_{\rm eff}(r_++\epsilon)=\frac{L^2-2 L \sqrt{L^2+4 m^2r_+^2}+4 m^2 r_+^2}{4 \text{rp}^2 \sqrt{L^2+4 m^2r_+^2}}+\frac{L}{4 r_+^3} \epsilon +{\cal O}(\epsilon^2).
	\end{align}
Solving $V'_{\rm eff}(r_++\epsilon)=0$ for small $\epsilon$, we find that
	\begin{align}
		L=\frac{2m}{\sqrt{3}}\left(r_++\frac{2}{3}\epsilon\right)+{\cal O}(\epsilon^2).
	\label{eq:near_horizon_exKerr}
	\end{align}
To distinguish between the local maximum and minimum, we calculate the second derivative of the effective potential \eqref{eq:second_derivative_eff_potential}.
This gives
	\begin{align}
		V''_{\rm eff}(r_++\epsilon)=\frac{m}{2\sqrt{3}r_+^2}+{\cal O}(\epsilon)>0,
	\end{align}
so that the solution \eqref{eq:near_extremal} corresponds to the local minimum.
Thus there are no near-horizon limits of local maxima.

\subsubsection{The non-extremal cases}
\begin{figure}[h]
	\centering
		\includegraphics[scale=0.5]{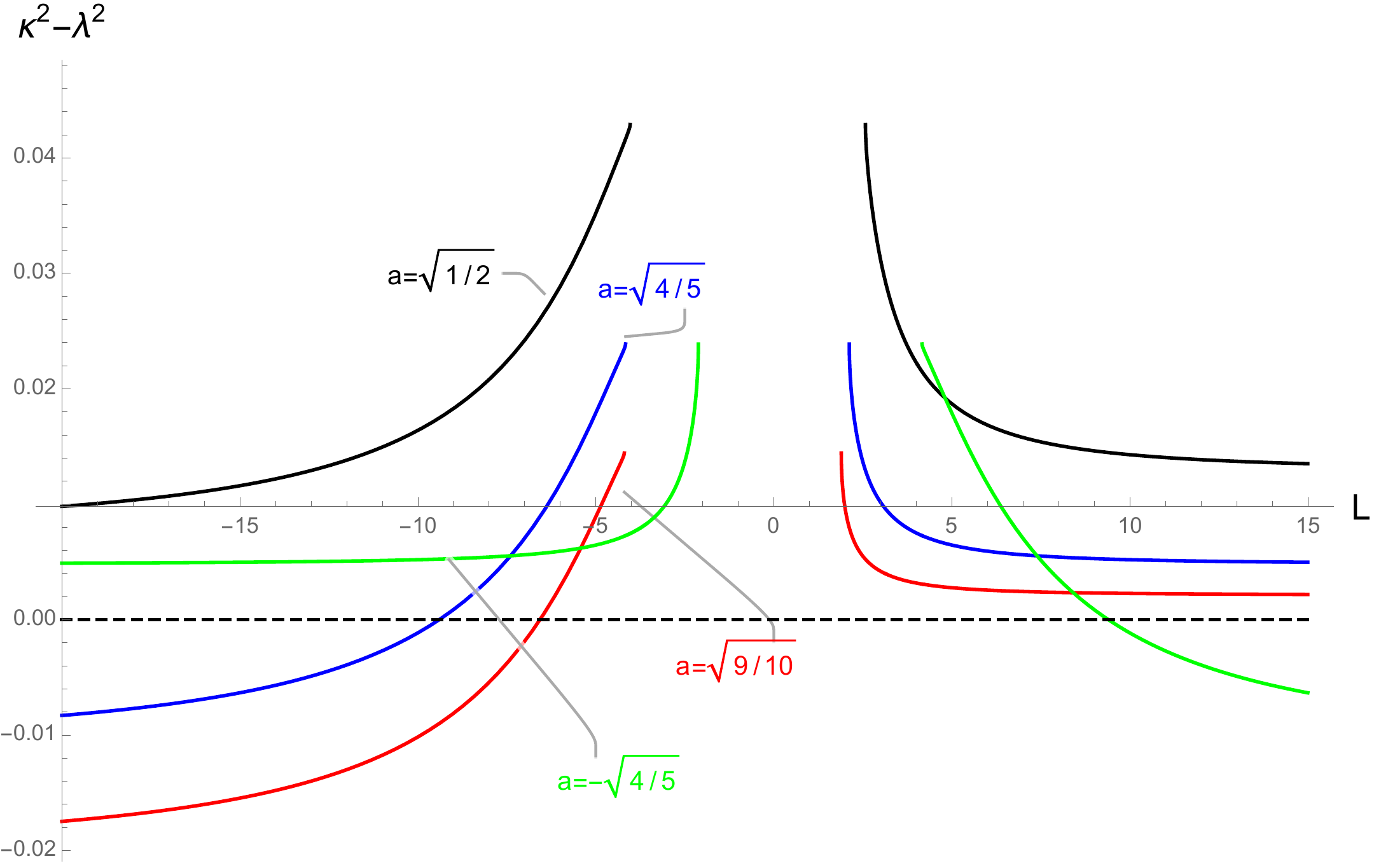}
		\caption{Numerical analysis of $\kappa^2-\lambda^2$ for the non-extremal Kerr black hole. The parameters corresponding to each color are as follows: The black lines represent $a=\sqrt{1/2}$. The blue lines represent $a=\sqrt{4/5}$. The red lines represent $a=\sqrt{9/10}$. The green lines represent $a=-\sqrt{4/5}$. The broken line represents the bound. We find the violation of the bound for $a=-\sqrt{4/5},\, \sqrt{4/5},\,$ and $\sqrt{9/10}$.  \label{fig:Lyapunov_nonexKerr}}
	\end{figure}

\noindent
We numerically analyze the bound for the non-extremal Kerr black hole.
Unlike the extremal case, the non-extremal Kerr black hole has a finite Hawking temperature.
The bound is given by
	\begin{align}
		\lambda^2\le\kappa^2=\frac{(r_+-r_-)^2}{4r_+^2(r_++r_-)^2}.
	\end{align}
The numerical results for $a=-\sqrt{4/5},\, \sqrt{1/2},\, \sqrt{4/5},\,$ and $\sqrt{9/10}$ are shown in Fig. \ref{fig:Lyapunov_nonexKerr}.
We see the violation of the bound for $a=-\sqrt{4/5},\, \sqrt{4/5},\,$ and $\sqrt{9/10}$.
We find the violation when the parameters of the black hole are close to extremal, which is the same as in the case of the non-extremal KN black hole.

Let us investigate the near-horizon region.
As before, we expand $V'_{\rm eff}(r_++\epsilon)$ around $\epsilon=0$:
	\begin{align}
		V'_{\rm eff}(r_++\epsilon)=\frac{\sqrt{(r_+-r_-) \left(L^2+m^2 (r_++r_-)^2\right)}}{2(r_++r_-)^2}\epsilon^{-1/2}-\frac{L r_-(3r_++r_-)}{\sqrt{r_+r_-} (r_++r_-)^3}+{\cal O}(\epsilon^{1/2}).
	\end{align}
For a small $\epsilon$, we can easily see that there are no real solutions to $V'_{\rm eff}(r_++\epsilon)=0$ when $L$ is real.
Therefore, the local extrema for the non-extremal Kerr black hole cannot approach the event horizon $r_+$.

\subsection{$J\to 0$ limit: The Reissner-Nordstr\"om black hole}
\noindent
We investigate the bound for the RN black hole.
For the RN black hole, the effective potential becomes simple; therefore, we can algebraically analyze the bound of the Lyapunov exponent.
The bound for the RN black hole without the angular momentum of the probe particle was studied in \cite{Hashimoto:2016dfz,Zhao:2018wkl,Lei:2020clg}.
Here we include the effect of the angular momentum $L$.

\subsubsection{The extremal cases}
\noindent 
As examined in \cite{Hashimoto:2016dfz}, the effective potential for the extremal RN black hole without the angular momentum of the particle does not have local maxima.
However, we show here that the effective potential with the angular momentum has a local maximum.

For the extremal RN black hole, we can suppose that $Q=r_+$ without loss of generality because the effective potential has only the combination $qQ$.
According to $V_{\rm eff}'(r_0)=0$, the position of the local extremum of the effective potential satisfies
	\begin{align}
		q=\frac{L^2(2r_+-r_0)+m^2r_+r_0^2}{r_+r_0\sqrt{L^2+m^2r_0^2}},
	\label{eq:extremum_exRN}
	\end{align}
where $r_0$ is the position of the local extremum.
Because the square of the Lyapunov exponent is proportional to $V_{\rm eff}''(r_0)$, the local extremum is the local maximum when $\lambda^2<0$.
When $L=0$, the dependence of $r_0$ is removed from \eqref{eq:extremum_exRN}, which implies the non-existence of extrema.
In other words, the local extremum is located at infinity.

We also consider the position of the local extremum for large $L$.
In this limit, we have 
	\begin{align}
		\frac{q}{L}=\frac{2r_+-r_0}{r_+r_0}.
	\end{align}
Thus we obtain $r_0=2r_+$ for $L\to\infty$.

The Lyapunov exponent defined in \eqref{eq:Lyapunov_exponent} is given by
	\begin{align}
		\lambda^2=\frac{L^2(r_0-r_+)^3\left(m^2r_0^2(3r_+-r_0)+2L^2r_+\right)}{r_0^6(L^2+m^2r_0^2)^2}.
	\label{eq:Lyapunov_exRN}
	\end{align}
From \eqref{eq:Lyapunov_exRN}, we see that the bound $\lambda^2\le 0$ is violated when 
	\begin{align}
		L\neq0 \quad{\rm for}\quad r_+<r_0\le 3r_+
	\label{eq:violation_condition_exRN_1}
	\end{align}
or
	\begin{align}
		L^2>\frac{m^2 r_0^2(r_0-3r_+)}{2r_+}\quad {\rm for}\quad r_0>3r_+.
	\label{eq:violation_condition_exRN_2}
	\end{align}
For the extremal RN black hole, we also observe the violation.
In this parameter region, the local maximum exists, which implies the violation.
For the other parameter region, there are no local maxima.

Next, we consider the near-horizon limit of the position of the local maximum.
To find the near-horizon limit, assuming $r_0=r_++\epsilon$, we expand \eqref{eq:extremum_exRN} around $\epsilon=0$:
	\begin{align}
		q=\frac{\sqrt{L^2+m^2 r_+^2}}{Q}-\frac{2 L^2}{Qr_+ \sqrt{L^2+m^2r_+^2}} \epsilon +{\cal O}(\epsilon^2).
	\end{align}
This implies that when the product of the charges $q$ and $Q$ is slightly smaller than $qQ=\sqrt{L^2+m^2r_+^2}$, the local maximum is placed near the event horizon.
Eq. \eqref{eq:violation_condition_exRN_1} implies that the bound is locally violated if $L\neq0$.

\subsubsection{The non-extremal cases}
\noindent 
The position of the local extremum for the non-extremal RN black hole obeys
	\begin{align}
		q=\frac{L^2 \left(-2r_0^2+3r_0(r_-+r_+)-4r_- r_+\right)+m^2 r_0^2 \left(r_0 (r_-+r_+)-2r_-r_+\right)}{2Qr_0\sqrt{(r_0-r_-) (r_0-r_+)\left(L^2+m^2 r_0^2\right)}},
	\label{eq:local_extremum_nonexRN}
	\end{align}
where $r_0$ is the position of the local extremum.
In the large-$L$ limit, we obtain
	\begin{align}
		\frac{q}{L}=\frac{-2r_0^2+3r_0(r_-+r_+)-4r_- r_+}{2 Q \sqrt{(r_0-r_-) (r_0-r_+)}}.
	\end{align}
Thus we find 
	\begin{align}
		r_0=\frac{1}{4}\left({3(r_-+r_+)}+\sqrt{9r_-^2-14r_-r_++9r_+^2}\right)
	\end{align}
for $L\to\infty$.
In particular, we see that $r_0=3r_+/2$ in the limit of the Schwarzschild black hole, $r_-\to0$, which coincides with Eq. (47) in \cite{Hashimoto:2016dfz}.

$\kappa^2-\lambda^2$ of the non-extremal RN black hole is given by
	\begin{align}
		\kappa^2-\lambda^2=&\frac{L^4 u+2 L^2 m^2 r_0^2 v+m^4 r_0^6 \left(r_0^4-r_+^4\right) (r_+-r_-)^2}{4 r_0^6 r_+^4\left(L^2+m^2 r_0^2\right)^2},
	\label{eq:Lyapunov_nonexRN}
	\end{align}
where we define $u$ and $v$ as follows:
	\begin{gather}
		u=r_0^6 (r_+-r_-)^2-4r_0^3r_+^4 (r_-+r_+)+3r_0^2r_+^4\left(r_-^2+6r_-r_++r_+^2\right)-12r_0r_- r_+^5 (r_-+r_+)+8r_-^2r_+^6, \\
		v=r_0^6(r_+-r_-)^2+2 r_0^4 r_+^4-6 r_0^3r_+^4 (r_-+r_+)+3 r_0^2 r_+^4 \left(r_-^2+6 r_-r_++r_+^2\right)-10 r_0 r_- r_+^5 (r_-+r_+)+6 r_-^2r_+^6.
	\end{gather}
To investigate the bound, we introduce a positive parameter $\delta$ defined by $M^2=Q^2+\delta^2$.
Using this parameter, the Cauchy horizon $r_-$ is represented as
	\begin{align}
		r_-=r_+-2\delta.
	\label{eq:near_extremal}
	\end{align}
The limit $\delta\to 0$ corresponds to the extremal limit, and the limit $\delta\to r_+/2$ corresponds to the Schwarzschild black hole.

Let us consider the near-extremal region.
Inserting \eqref{eq:near_extremal} into \eqref{eq:Lyapunov_nonexRN}, we expand around $\delta=0$:
	\begin{align}
		\kappa^2-\lambda^2=&{-\frac{L^2(r_0-r_+)^3\left(m^2r_0^2(3r_+-r_0)+2L^2r_+\right)}{r_0^6(L^2+m^2r_0^2)^2}} \nonumber \\
		&\quad+\frac{2 L^2 (r_0-r_+)^2 \left(L^2 (r_0-4r_+)+3 m^2 r_0^2 (r_0-2 r_+)\right)}{r_0^6 \left(L^2+m^2r_0^2\right)^2}\delta +{\cal O}(\delta^2).
	\label{eq:near_extremal_limit_nonexRN}
	\end{align}
The leading terms coincide with \eqref{eq:Lyapunov_exRN}.
Thus, the bound for the near-extremal RN black hole is violated when \eqref{eq:violation_condition_exRN_1} or \eqref{eq:violation_condition_exRN_2} is satisfied.

We also consider the Schwarzschild limit.
Inserting \eqref{eq:near_extremal} into \eqref{eq:Lyapunov_nonexRN}, we expand around $\delta=r_+/2$,
	\begin{align}
		\kappa^2-\lambda^2=&\frac{\mu}{r_0^4r_+^2 \left(L^2+m^2r_0^2\right)^2} +\frac{\nu}{r_0^5r_+^3\left(L^2+m^2 r_0^2\right)^2}(\delta-r_+/2)+{\cal O}\left((\delta-r_+/2)^2\right),
	\end{align}
where we define
	\begin{gather}
		\mu=L^4 \left(r_0^4-4 r_0r_+^3+3r_+^4\right)+2 L^2 m^2 r_0^2 \left(r_0^4+2 r_0^2r_+^2-6 r_0r_+^3+3r_+^4\right)+m^4r_0^4 \left(r_0^4-r_+^4\right), \\
		\nu=L^4 \left(r_0^5+2 r_0^2r_+^3-9 r_0r_+^4+6r_+^5\right)+2 L^2 m^2 r_0^2 \left(r_0^5+3 r_0^2r_+^3-9 r_0r_+^4+5r_+^5\right)+m^4r_0^5 \left(r_0^4-r_+^4\right).
	\end{gather}
We can easily show that $\mu>0$.
Therefore the bound is satisfied for the near-Schwarzschild region.
We conclude that the bound for the non-extremal RN black hole is violated when the parameters of the black hole approach the extreme.

We consider the near-horizon.
Similarly, according to \eqref{eq:local_extremum_nonexRN}, we find
	\begin{align}
		q=\frac{\sqrt{(r_+-r_-) \left(L^2+m^2 r_+^2\right)}}{2 Q }\epsilon^{-1/2}+\frac{\left(L^2 (8 r_--5 r_+)+m^2r_+^2 (2r_-+r_+)\right)}{4 Qr_+\sqrt{(r_+-r_-) \left(L^2+m^2 r_+^2\right)}}\epsilon^{1/2}.
	\end{align}
In the limit $q\to\infty$, the local maximum is close to the horizon, where the angular momentum $L$ is finite.

We analyze the bound for the near-horizon region using \eqref{eq:Lyapunov_nonexRN},
	\begin{align}
		\kappa^2-\lambda^2=\frac{m^2(r_+-r_-)^2}{r_+^3 \left(L^2+m^2r_+^2\right)} \epsilon  +{\cal O}(\epsilon^2).
	\end{align}
Because the coefficient of the leading term is positive, the bound for the near-horizon is satisfied even if $L\neq0$.

Can we find a modified bound that is not broken by the angular momenta?
A simple possibility of the modification is $\lambda\le C_0\kappa$, where for simplicity, we assume $C_0$ is a parameter-independent constant.
However, this modification does not work.
To see this, we focus on the near-extremal region since in this region.
Then the modification becomes $\lambda\le C_0\kappa\sim C_0\delta$.
As we have seen, e.g. Eq. \eqref{eq:near_extremal_limit_nonexRN}, the leading term of the Lyapunov exponent is ${\cal O}(\delta^0)$.
Thus the modification is negligible, and the modified bound can be violated in the near-extremal region.
The another possibility of the modification is $\lambda\le \kappa+C_1$, where $C_1$ is also a parameter-independent constant.
This modification does not work again: for example, in the case of the extremal RN black hole \eqref{eq:Lyapunov_exRN} is roughly given by $\lambda\sim L^2$ for large $L$, which implies that the modified bound, $\lambda\le \kappa+C_1$, is violated for large $L$.

In this section, we analyzed the bound on the Lyapunov exponent.
The results of this analysis are summarized in Table \ref{table:summary}.
When the position of the local maximum is located at a general point, depending on the parameters, the bound can be violated.
When the position of the local maximum is located near the event horizon, we can analyze the bound algebraically.
In this situation, the KN, Kerr, and RN black holes show quite different results from each other.

\section{Conclusions}
\label{sec:conclusions}
\noindent 
We investigated the Lyapunov exponent for a probe particle around a KN black hole.
The effect of the angular momentum of the particle was considered in the equation of motion.
With the static gauge, we calculated the effective potential.
The Lyapunov exponent of the particle is conjectured to be upper bounded by the surface gravity \cite{Maldacena:2015waa}.
Applying the bound on the Lyapunov exponent to the motion of the particle, we found violation of the bound in several cases of the KN black hole: the Kerr black hole, the RN black hole, and the near-horizon limit.
Our results are summarized in Table \ref{table:summary}.
	\begin{table}[h]
	\centering
		\begin{tabular}{c|ccc}
			Bound & Far from the horizon & Near-horizon & Event horizon\\
			\hline 
			Ex. KN & $\times$ & $\times$ & $\bigcirc$ \\
			Non-ex. KN & $\times$ & $\bigcirc$ & $\bigcirc$ \\
			Ex. Kerr & $\times$ & $-$ & $-$ \\
			Non-ex. Kerr & $\times$ & $-$ & $-$\\
			Ex. RN & $\times$ & $\times$ & $\bigcirc$ \\
			Non-ex. RN & $\times$ & $\bigcirc$ & $\bigcirc$
		\end{tabular}
		\caption{Summary of results.``Far from the horizon'' means that the local maximum is located at a general point. ``Near-horizon'' means that the local maximum is located near the event horizon. ``Event horizon'' means that the local maximum is located at the event horizon. ``$\bigcirc$'' represents that the bound is satisfied for any parameter regions of the black hole and particle. ``$\times$'' represents that we found a violation of the upper bound of the Lyapunov exponent that depends on the parameters. ``$-$'' represents that the local maximum cannot be close to the near-horizon.\label{table:summary}}
	\end{table}

In the motion of the particle, the effective potential near the local maximum can be approximated by the inverse harmonic oscillator.
The inverse harmonic oscillator exhibits chaos, and the Lyapunov exponent of the system is proportional to the second derivative of the effective potential at the local maximum.
Thus, the Lyapunov exponent depends on the parameters, such as the charges and angular momenta of the black hole and the particle.
In particular, we found that the behavior of the Lyapunov exponent changed significantly owing to the contribution of the angular momenta.
In this study, the behaviors are categorized into three cases depending on the type of black hole: the KN black hole, which has a finite charge and angular momentum; the Kerr black hole, which is the $Q\to 0$ limit of the KN black hole; and the RN black hole, which is the $a\to 0$ limit of the KN black hole, where $Q$ and $a$ are the charge and angular momentum of the black hole, respectively.
For the general cases in which the local maximum is located at an arbitrary place, we investigated the behaviors of the Lyapunov exponent with various parameter sets of the black hole and the particle.
The location of the local maximum is closely associated with the Lyapunov exponent.
Because the effective potential depends on the parameters, the existence of the local maximum depends on the parameters. 
We found that a local maximum can exist (and the bound can be violated) for particular sets of parameters.
For the near-horizon cases where the local maximum is located near the event horizon, the results are remarkable.
In particular, we can perform algebraic analysis for these cases.
In the case of the KN black hole, the bound can be violated because the second derivative of the effective potential can be positive.
In contrast, for the cases of the Kerr and RN black holes, the bound is satisfied except for the extremal RN black hole.
For the extremal RN black hole, the bound is violated if and only if the angular momentum of the particle is finite.
Therefore, we see that the angular momenta of the black hole and particle play an important role in exceeding the bound.
When the local maximum is exactly located at the event horizon, the results change.
For the KN and RN black holes, the Lyapunov exponent coincides with the surface gravity.
For the Kerr black hole, the local maximum cannot be located at the event horizon.

In this study, we focus on the KN black hole, which is an asymptotically flat black hole without the cosmological constant.
It is natural to wonder how angular momentum affects the bound on the Lyapunov exponent when the cosmological constant is nonzero.
In particular, if the cosmological constant is negative, the relationship between the Lyapunov exponent and its bound can be studied in the AdS/CFT correspondence. 

\section*{Acknowledgements}
\noindent This work was supported by the National Research Foundation of Korea (NRF) grant funded by the Korea government (MSIT) (NRF-2018R1C1B6004349) and the Dongguk University Research Fund of 2021. BG appreciates APCTP for its hospitality during the topical research program, {\it Gravity and Cosmology}.

\bibliographystyle{jhep}
\bibliography{ref}

\providecommand{\href}[2]{#2}\begingroup\raggedright\begin{thebibliography}{10}

\bibitem{Maldacena:1997re}
J.~M. Maldacena, \emph{{The Large N limit of superconformal field theories and
  supergravity}}, \href{https://doi.org/10.1023/A:1026654312961}{\emph{Adv.
  Theor. Math. Phys.} {\bfseries 2} (1998) 231}
  [\href{https://arxiv.org/abs/hep-th/9711200}{{\ttfamily hep-th/9711200}}].

\bibitem{Gubser:1998bc}
S.~S. Gubser, I.~R. Klebanov and A.~M. Polyakov, \emph{{Gauge theory
  correlators from noncritical string theory}},
  \href{https://doi.org/10.1016/S0370-2693(98)00377-3}{\emph{Phys. Lett. B}
  {\bfseries 428} (1998) 105}
  [\href{https://arxiv.org/abs/hep-th/9802109}{{\ttfamily hep-th/9802109}}].

\bibitem{Witten:1998qj}
E.~Witten, \emph{{Anti-de Sitter space and holography}},
  \href{https://doi.org/10.4310/ATMP.1998.v2.n2.a2}{\emph{Adv. Theor. Math.
  Phys.} {\bfseries 2} (1998) 253}
  [\href{https://arxiv.org/abs/hep-th/9802150}{{\ttfamily hep-th/9802150}}].

\bibitem{Witten:1998zw}
E.~Witten, \emph{{Anti-de Sitter space, thermal phase transition, and
  confinement in gauge theories}},
  \href{https://doi.org/10.4310/ATMP.1998.v2.n3.a3}{\emph{Adv. Theor. Math.
  Phys.} {\bfseries 2} (1998) 505}
  [\href{https://arxiv.org/abs/hep-th/9803131}{{\ttfamily hep-th/9803131}}].

\bibitem{Larkin:1969qu}
A.~I. Larkin and Y.~N. Ovchinnikov, \emph{{Quasiclassical method in the theory
  of super- conductivity}}, {\emph{JETP} {\bfseries 28} (1969) 1200}.

\bibitem{Maldacena:2015waa}
J.~Maldacena, S.~H. Shenker and D.~Stanford, \emph{{A bound on chaos}},
  \href{https://doi.org/10.1007/JHEP08(2016)106}{\emph{JHEP} {\bfseries 08}
  (2016) 106} [\href{https://arxiv.org/abs/1503.01409}{{\ttfamily
  1503.01409}}].

\bibitem{Shenker:2013pqa}
S.~H. Shenker and D.~Stanford, \emph{{Black holes and the butterfly effect}},
  \href{https://doi.org/10.1007/JHEP03(2014)067}{\emph{JHEP} {\bfseries 03}
  (2014) 067} [\href{https://arxiv.org/abs/1306.0622}{{\ttfamily 1306.0622}}].

\bibitem{Shenker:2013yza}
S.~H. Shenker and D.~Stanford, \emph{{Multiple Shocks}},
  \href{https://doi.org/10.1007/JHEP12(2014)046}{\emph{JHEP} {\bfseries 12}
  (2014) 046} [\href{https://arxiv.org/abs/1312.3296}{{\ttfamily 1312.3296}}].

\bibitem{Sachdev:1992fk}
S.~Sachdev and J.~Ye, \emph{{Gapless spin fluid ground state in a random,
  quantum Heisenberg magnet}},
  \href{https://doi.org/10.1103/PhysRevLett.70.3339}{\emph{Phys. Rev. Lett.}
  {\bfseries 70} (1993) 3339}
  [\href{https://arxiv.org/abs/cond-mat/9212030}{{\ttfamily
  cond-mat/9212030}}].

\bibitem{Kitaev:2014hi}
A.~Kitaev, \emph{Hidden correlations in the hawking radiation and thermal
  noise}, {\emph{talk given at Fundamental Physics Prize Symposium} (Nov. 2014)
  }.

\bibitem{Polchinski:2016xgd}
J.~Polchinski and V.~Rosenhaus, \emph{{The Spectrum in the Sachdev-Ye-Kitaev
  Model}}, \href{https://doi.org/10.1007/JHEP04(2016)001}{\emph{JHEP}
  {\bfseries 04} (2016) 001}
  [\href{https://arxiv.org/abs/1601.06768}{{\ttfamily 1601.06768}}].

\bibitem{Maldacena:2016hyu}
J.~Maldacena and D.~Stanford, \emph{{Remarks on the Sachdev-Ye-Kitaev model}},
  \href{https://doi.org/10.1103/PhysRevD.94.106002}{\emph{Phys. Rev. D}
  {\bfseries 94} (2016) 106002}
  [\href{https://arxiv.org/abs/1604.07818}{{\ttfamily 1604.07818}}].

\bibitem{Witten:2016iux}
E.~Witten, \emph{{An SYK-Like Model Without Disorder}},
  \href{https://doi.org/10.1088/1751-8121/ab3752}{\emph{J. Phys. A} {\bfseries
  52} (2019) 474002} [\href{https://arxiv.org/abs/1610.09758}{{\ttfamily
  1610.09758}}].

\bibitem{Gross:2016kjj}
D.~J. Gross and V.~Rosenhaus, \emph{{A Generalization of Sachdev-Ye-Kitaev}},
  \href{https://doi.org/10.1007/JHEP02(2017)093}{\emph{JHEP} {\bfseries 02}
  (2017) 093} [\href{https://arxiv.org/abs/1610.01569}{{\ttfamily
  1610.01569}}].

\bibitem{Kitaev:2015as}
A.~Kitaev, \emph{A simple model of quantum holography}, {\emph{talks given at
  KITP} (April and May 2015) }.

\bibitem{Maldacena:2016upp}
J.~Maldacena, D.~Stanford and Z.~Yang, \emph{{Conformal symmetry and its
  breaking in two dimensional Nearly Anti-de-Sitter space}},
  \href{https://doi.org/10.1093/ptep/ptw124}{\emph{PTEP} {\bfseries 2016}
  (2016) 12C104} [\href{https://arxiv.org/abs/1606.01857}{{\ttfamily
  1606.01857}}].

\bibitem{Kitaev:2017awl}
A.~Kitaev and S.~J. Suh, \emph{{The soft mode in the Sachdev-Ye-Kitaev model
  and its gravity dual}},
  \href{https://doi.org/10.1007/JHEP05(2018)183}{\emph{JHEP} {\bfseries 05}
  (2018) 183} [\href{https://arxiv.org/abs/1711.08467}{{\ttfamily
  1711.08467}}].

\bibitem{Jackiw:1984je}
R.~Jackiw, \emph{{Lower Dimensional Gravity}},
  \href{https://doi.org/10.1016/0550-3213(85)90448-1}{\emph{Nucl. Phys. B}
  {\bfseries 252} (1985) 343}.

\bibitem{Teitelboim:1983ux}
C.~Teitelboim, \emph{{Gravitation and Hamiltonian Structure in Two Space-Time
  Dimensions}}, \href{https://doi.org/10.1016/0370-2693(83)90012-6}{\emph{Phys.
  Lett. B} {\bfseries 126} (1983) 41}.

\bibitem{Dettmann:1994dj}
C.~P. Dettmann, N.~E. Frankel and N.~J. Cornish, \emph{{Fractal basins and
  chaotic trajectories in multi - black hole space-times}},
  \href{https://doi.org/10.1103/PhysRevD.50.R618}{\emph{Phys. Rev. D}
  {\bfseries 50} (1994) R618}
  [\href{https://arxiv.org/abs/gr-qc/9402027}{{\ttfamily gr-qc/9402027}}].

\bibitem{Suzuki:1996gm}
S.~Suzuki and K.-i. Maeda, \emph{{Chaos in Schwarzschild space-time: The motion
  of a spinning particle}},
  \href{https://doi.org/10.1103/PhysRevD.55.4848}{\emph{Phys. Rev. D}
  {\bfseries 55} (1997) 4848}
  [\href{https://arxiv.org/abs/gr-qc/9604020}{{\ttfamily gr-qc/9604020}}].

\bibitem{Suzuki:1999si}
S.~Suzuki and K.-i. Maeda, \emph{{Signature of chaos in gravitational waves
  from a spinning particle}},
  \href{https://doi.org/10.1103/PhysRevD.61.024005}{\emph{Phys. Rev. D}
  {\bfseries 61} (2000) 024005}
  [\href{https://arxiv.org/abs/gr-qc/9910064}{{\ttfamily gr-qc/9910064}}].

\bibitem{Dalui:2018qqv}
S.~Dalui, B.~R. Majhi and P.~Mishra, \emph{{Presence of horizon makes particle
  motion chaotic}},
  \href{https://doi.org/10.1016/j.physletb.2018.11.050}{\emph{Phys. Lett. B}
  {\bfseries 788} (2019) 486}
  [\href{https://arxiv.org/abs/1803.06527}{{\ttfamily 1803.06527}}].

\bibitem{Li:2018wtz}
D.~Li and X.~Wu, \emph{{Chaotic motion of neutral and charged particles in a
  magnetized Ernst-Schwarzschild spacetime}},
  \href{https://doi.org/10.1140/epjp/i2019-12502-9}{\emph{Eur. Phys. J. Plus}
  {\bfseries 134} (2019) 96}
  [\href{https://arxiv.org/abs/1803.02119}{{\ttfamily 1803.02119}}].

\bibitem{Lukes-Gerakopoulos:2016xoc}
G.~Lukes-Gerakopoulos, \emph{{Comment on ''Chaotic orbits for spinning
  particles in Schwarzschild spacetime''}},
  \href{https://doi.org/10.1103/PhysRevD.94.108501}{\emph{Phys. Rev. D}
  {\bfseries 94} (2016) 108501}
  [\href{https://arxiv.org/abs/1604.02955}{{\ttfamily 1604.02955}}].

\bibitem{Verhaaren:2009md}
C.~Verhaaren and E.~W. Hirschmann, \emph{{Chaotic orbits for spinning particles
  in Schwarzschild spacetime}},
  \href{https://doi.org/10.1103/PhysRevD.81.124034}{\emph{Phys. Rev. D}
  {\bfseries 81} (2010) 124034}
  [\href{https://arxiv.org/abs/0912.0031}{{\ttfamily 0912.0031}}].

\bibitem{Han:2008zzf}
W.~Han, \emph{{Chaos and dynamics of spinning particles in Kerr spacetime}},
  \href{https://doi.org/10.1007/s10714-007-0598-9}{\emph{Gen. Rel. Grav.}
  {\bfseries 40} (2008) 1831}
  [\href{https://arxiv.org/abs/1006.2229}{{\ttfamily 1006.2229}}].

\bibitem{Cornish:2001jy}
N.~J. Cornish, \emph{{Chaos and gravitational waves}},
  \href{https://doi.org/10.1103/PhysRevD.64.084011}{\emph{Phys. Rev. D}
  {\bfseries 64} (2001) 084011}
  [\href{https://arxiv.org/abs/gr-qc/0106062}{{\ttfamily gr-qc/0106062}}].

\bibitem{Cardoso:2008bp}
V.~Cardoso, A.~S. Miranda, E.~Berti, H.~Witek and V.~T. Zanchin,
  \emph{{Geodesic stability, Lyapunov exponents and quasinormal modes}},
  \href{https://doi.org/10.1103/PhysRevD.79.064016}{\emph{Phys. Rev. D}
  {\bfseries 79} (2009) 064016}
  [\href{https://arxiv.org/abs/0812.1806}{{\ttfamily 0812.1806}}].

\bibitem{Pradhan:2012rkk}
P.~Pradhan, \emph{{Stability analysis and quasinormal modes of
  Reissner\textendash{}Nordstr\o{}m space-time via Lyapunov exponent}},
  \href{https://doi.org/10.1007/s12043-016-1214-x}{\emph{Pramana} {\bfseries
  87} (2016) 5} [\href{https://arxiv.org/abs/1205.5656}{{\ttfamily
  1205.5656}}].

\bibitem{Pradhan:2012qf}
P.~P. Pradhan, \emph{{ISCO, Lyapunov exponent and Kolmogorov-Sinai entropy for
  Kerr-Newman Black hole}},  \href{https://arxiv.org/abs/1212.5758}{{\ttfamily
  1212.5758}}.

\bibitem{Pradhan:2013bli}
P.~P. Pradhan, \emph{{Lyapunov Exponent and Charged Myers Perry Spacetimes}},
  \href{https://doi.org/10.1140/epjc/s10052-013-2477-8}{\emph{Eur. Phys. J. C}
  {\bfseries 73} (2013) 2477}
  [\href{https://arxiv.org/abs/1302.2536}{{\ttfamily 1302.2536}}].

\bibitem{Pradhan:2014tva}
P.~Pradhan, \emph{{Circular Geodesics in Tidal Charged Black Hole}},
  \href{https://doi.org/10.1142/S0219887818500111}{\emph{Int. J. Geom. Meth.
  Mod. Phys.} {\bfseries 15} (2017) 1850011}
  [\href{https://arxiv.org/abs/1412.8123}{{\ttfamily 1412.8123}}].

\bibitem{Jawad:2016kgt}
A.~Jawad, F.~Ali, M.~U. Shahzad and G.~Abbas, \emph{{Dynamics of particles
  around time conformal Schwarzschild black hole}},
  \href{https://doi.org/10.1140/epjc/s10052-016-4422-0}{\emph{Eur. Phys. J. C}
  {\bfseries 76} (2016) 586}
  [\href{https://arxiv.org/abs/1610.05610}{{\ttfamily 1610.05610}}].

\bibitem{Jai-akson:2017ldo}
P.~Jai-akson, A.~Chatrabhuti, O.~Evnin and L.~Lehner, \emph{{Black hole merger
  estimates in Einstein-Maxwell and Einstein-Maxwell-dilaton gravity}},
  \href{https://doi.org/10.1103/PhysRevD.96.044031}{\emph{Phys. Rev. D}
  {\bfseries 96} (2017) 044031}
  [\href{https://arxiv.org/abs/1706.06519}{{\ttfamily 1706.06519}}].

\bibitem{Chen:2016tmr}
S.~Chen, M.~Wang and J.~Jing, \emph{{Chaotic motion of particles in the
  accelerating and rotating black holes spacetime}},
  \href{https://doi.org/10.1007/JHEP09(2016)082}{\emph{JHEP} {\bfseries 09}
  (2016) 082} [\href{https://arxiv.org/abs/1604.02785}{{\ttfamily
  1604.02785}}].

\bibitem{Yurtsever:1994yb}
U.~Yurtsever, \emph{{Geometry of chaos in the two center problem in general
  relativity}}, \href{https://doi.org/10.1103/PhysRevD.52.3176}{\emph{Phys.
  Rev. D} {\bfseries 52} (1995) 3176}
  [\href{https://arxiv.org/abs/gr-qc/9412031}{{\ttfamily gr-qc/9412031}}].

\bibitem{Barrow:1981sx}
J.~D. Barrow, \emph{{Chaotic behavior in general relativity}},
  \href{https://doi.org/10.1016/0370-1573(82)90171-5}{\emph{Phys. Rept.}
  {\bfseries 85} (1982) 1}.

\bibitem{Bombelli:1991eg}
L.~Bombelli and E.~Calzetta, \emph{{Chaos around a black hole}},
  \href{https://doi.org/10.1088/0264-9381/9/12/004}{\emph{Class. Quant. Grav.}
  {\bfseries 9} (1992) 2573}.

\bibitem{Letelier:1997uv}
P.~S. Letelier and W.~M. Vieira, \emph{{Chaos and rotating black holes with
  halos}}, \href{https://doi.org/10.1103/PhysRevD.56.8095}{\emph{Phys. Rev. D}
  {\bfseries 56} (1997) 8095}
  [\href{https://arxiv.org/abs/gr-qc/9712008}{{\ttfamily gr-qc/9712008}}].

\bibitem{deMoura:1999wf}
A.~P.~S. de~Moura and P.~S. Letelier, \emph{{Chaos and fractals in geodesic
  motions around a nonrotating black hole with an external halo}},
  \href{https://doi.org/10.1103/PhysRevE.61.6506}{\emph{Phys. Rev. E}
  {\bfseries 61} (2000) 6506}
  [\href{https://arxiv.org/abs/chao-dyn/9910035}{{\ttfamily
  chao-dyn/9910035}}].

\bibitem{Yi:2020shw}
M.~Yi and X.~Wu, \emph{{Dynamics of charged particles around a magnetically
  deformed Schwarzschild black hole}},
  \href{https://doi.org/10.1088/1402-4896/aba4c2}{\emph{Phys. Scripta}
  {\bfseries 95} (2020) 085008}.

\bibitem{Zelenka:2019nyp}
O.~Zelenka, G.~Lukes-Gerakopoulos, V.~Witzany and O.~Kop\'a\v{c}ek,
  \emph{{Growth of resonances and chaos for a spinning test particle in the
  Schwarzschild background}},
  \href{https://doi.org/10.1103/PhysRevD.101.024037}{\emph{Phys. Rev. D}
  {\bfseries 101} (2020) 024037}
  [\href{https://arxiv.org/abs/1911.00414}{{\ttfamily 1911.00414}}].

\bibitem{Mondal:2021exj}
M.~Mondal, F.~Rahaman and K.~N. Singh, \emph{{Lyapunov exponent ISCO and
  Kolmogorov Senai entropy for Kerr Kiselev black hole}},
  \href{https://doi.org/10.1140/epjc/s10052-021-08888-1}{\emph{Eur. Phys. J. C}
  {\bfseries 81} (2021) 84} [\href{https://arxiv.org/abs/2102.02667}{{\ttfamily
  2102.02667}}].

\bibitem{Lukes-Gerakopoulos:2016udm}
G.~Lukes-Gerakopoulos, \emph{{Spinning particles moving around black holes:
  integrability and chaos}},  in \emph{{14th Marcel Grossmann Meeting on Recent
  Developments in Theoretical and Experimental General Relativity,
  Astrophysics, and Relativistic Field Theories}}, vol.~2, pp.~1960--1965,
  2017, \href{https://arxiv.org/abs/1606.09430}{{\ttfamily 1606.09430}},
  \href{https://doi.org/10.1142/9789813226609_0209}{DOI}.

\bibitem{Setare:2010zd}
M.~R. Setare and D.~Momeni, \emph{{Geodesic stability for KS Black hole in
  Horava-Lifshitz gravity via Lyapunov exponents}},
  \href{https://doi.org/10.1007/s10773-010-0498-8}{\emph{Int. J. Theor. Phys.}
  {\bfseries 50} (2011) 106} [\href{https://arxiv.org/abs/1001.3767}{{\ttfamily
  1001.3767}}].

\bibitem{Liu:2017fjx}
C.-Y. Liu, D.-S. Lee and C.-Y. Lin, \emph{{Geodesic motion of neutral particles
  around a Kerr\textendash{}Newman black hole}},
  \href{https://doi.org/10.1088/1361-6382/aa903b}{\emph{Class. Quant. Grav.}
  {\bfseries 34} (2017) 235008}
  [\href{https://arxiv.org/abs/1706.05466}{{\ttfamily 1706.05466}}].

\bibitem{PandoZayas:2010xpn}
L.~A. Pando~Zayas and C.~A. Terrero-Escalante, \emph{{Chaos in the Gauge /
  Gravity Correspondence}},
  \href{https://doi.org/10.1007/JHEP09(2010)094}{\emph{JHEP} {\bfseries 09}
  (2010) 094} [\href{https://arxiv.org/abs/1007.0277}{{\ttfamily 1007.0277}}].

\bibitem{Basu:2012ae}
P.~Basu, D.~Das, A.~Ghosh and L.~A. Pando~Zayas, \emph{{Chaos around
  Holographic Regge Trajectories}},
  \href{https://doi.org/10.1007/JHEP05(2012)077}{\emph{JHEP} {\bfseries 05}
  (2012) 077} [\href{https://arxiv.org/abs/1201.5634}{{\ttfamily 1201.5634}}].

\bibitem{Hashimoto:2016wme}
K.~Hashimoto, K.~Murata and K.~Yoshida, \emph{{Chaos in chiral condensates in
  gauge theories}},
  \href{https://doi.org/10.1103/PhysRevLett.117.231602}{\emph{Phys. Rev. Lett.}
  {\bfseries 117} (2016) 231602}
  [\href{https://arxiv.org/abs/1605.08124}{{\ttfamily 1605.08124}}].

\bibitem{Nunez:2018ags}
C.~N\'u\~nez, J.~M. Pen\'\i{}n, D.~Roychowdhury and J.~Van~Gorsel, \emph{{The
  non-Integrability of Strings in Massive Type IIA and their Holographic
  duals}}, \href{https://doi.org/10.1007/JHEP06(2018)078}{\emph{JHEP}
  {\bfseries 06} (2018) 078}
  [\href{https://arxiv.org/abs/1802.04269}{{\ttfamily 1802.04269}}].

\bibitem{Hashimoto:2018fkb}
K.~Hashimoto, K.~Murata and N.~Tanahashi, \emph{{Chaos of Wilson Loop from
  String Motion near Black Hole Horizon}},
  \href{https://doi.org/10.1103/PhysRevD.98.086007}{\emph{Phys. Rev. D}
  {\bfseries 98} (2018) 086007}
  [\href{https://arxiv.org/abs/1803.06756}{{\ttfamily 1803.06756}}].

\bibitem{Akutagawa:2018yoe}
T.~Akutagawa, K.~Hashimoto, T.~Miyazaki and T.~Ota, \emph{{Phase diagram of QCD
  chaos in linear sigma models and holography}},
  \href{https://doi.org/10.1093/ptep/pty055}{\emph{PTEP} {\bfseries 2018}
  (2018) 063B01} [\href{https://arxiv.org/abs/1804.01737}{{\ttfamily
  1804.01737}}].

\bibitem{Akutagawa:2019awh}
T.~Akutagawa, K.~Hashimoto, K.~Murata and T.~Ota, \emph{{Chaos of QCD string
  from holography}},
  \href{https://doi.org/10.1103/PhysRevD.100.046009}{\emph{Phys. Rev. D}
  {\bfseries 100} (2019) 046009}
  [\href{https://arxiv.org/abs/1903.04718}{{\ttfamily 1903.04718}}].

\bibitem{Cubrovic:2019qee}
M.~\v{C}ubrovi\'c, \emph{{The bound on chaos for closed strings in Anti-de
  Sitter black hole backgrounds}},
  \href{https://doi.org/10.1007/JHEP12(2019)150}{\emph{JHEP} {\bfseries 12}
  (2019) 150} [\href{https://arxiv.org/abs/1904.06295}{{\ttfamily
  1904.06295}}].

\bibitem{Hashimoto:2016dfz}
K.~Hashimoto and N.~Tanahashi, \emph{{Universality in Chaos of Particle Motion
  near Black Hole Horizon}},
  \href{https://doi.org/10.1103/PhysRevD.95.024007}{\emph{Phys. Rev. D}
  {\bfseries 95} (2017) 024007}
  [\href{https://arxiv.org/abs/1610.06070}{{\ttfamily 1610.06070}}].

\bibitem{Morita:2021syq}
T.~Morita, \emph{{Extracting Classical Lyapunov Exponent from One-Dimensional
  Quantum Mechanics}},  \href{https://arxiv.org/abs/2105.09603}{{\ttfamily
  2105.09603}}.

\bibitem{Morita:2021mfi}
T.~Morita, \emph{{Analogous Hawking Radiation in Butterfly Effect}},  in
  \emph{{4th International Conference on Holography, String Theory and Discrete
  Approach in Hanoi, Vietnam}}, 1, 2021,
  \href{https://arxiv.org/abs/2101.02435}{{\ttfamily 2101.02435}}.

\bibitem{Bhattacharyya:2020art}
A.~Bhattacharyya, W.~Chemissany, S.~S. Haque, J.~Murugan and B.~Yan, \emph{{The
  Multi-faceted Inverted Harmonic Oscillator: Chaos and Complexity}},
  \href{https://doi.org/10.21468/SciPostPhysCore.4.1.002}{\emph{SciPost Phys.
  Core} {\bfseries 4} (2021) 002}
  [\href{https://arxiv.org/abs/2007.01232}{{\ttfamily 2007.01232}}].

\bibitem{Zhao:2018wkl}
Q.-Q. Zhao, Y.-Z. Li and H.~Lu, \emph{{Static Equilibria of Charged Particles
  Around Charged Black Holes: Chaos Bound and Its Violations}},
  \href{https://doi.org/10.1103/PhysRevD.98.124001}{\emph{Phys. Rev. D}
  {\bfseries 98} (2018) 124001}
  [\href{https://arxiv.org/abs/1809.04616}{{\ttfamily 1809.04616}}].

\bibitem{Lei:2020clg}
Y.-Q. Lei, X.-H. Ge and C.~Ran, \emph{{Chaos of particle motion near a black
  hole with quasitopological electromagnetism}},
  \href{https://doi.org/10.1103/PhysRevD.104.046020}{\emph{Phys. Rev. D}
  {\bfseries 104} (2021) 046020}
  [\href{https://arxiv.org/abs/2008.01384}{{\ttfamily 2008.01384}}].

\end{thebibliography}\endgroup
\end{document}